\documentclass[twocolumn,aps]{revtex4}
\def\[{\left\lbrack}
\def\]{\right\rbrack}

\def\({\left(}
\def\){\right)}
\newcommand{\be}{\begin{equation}}
\newcommand{\ee}{\end{equation}}
\newcommand{\ea}{\end{eqnarray}}
\newcommand{\ba}{\begin{eqnarray}}

\begin{document}

\title{A New Approach to Canonical Quantization of the Radiation Damping}

\author{A.C.R. Mendes, C. Neves, W. Oliveira and F.I. Takakura }
\thanks{\noindent e-mail:albert@fisica.ufjf.br, cneves@fisica.ufjf.br,\\ wilson@fisica.ufjf.br, takakura@fisica.ufjf.br}
\affiliation{Departamento de F\'{\i}sica, Universidade Federal de
Juiz de Fora, 36036-330, Juiz de Fora, MG, Brasil}

\begin{abstract}
Inspired in some works about quantization of dissipative
systems, in particular of the damped harmonic
oscillator\cite{MB,RB,12}, we consider the dissipative system of a
charge interacting with its own radiation, which originates the
radiation damping (RD). Using the indirect Lagrangian representation we
obtained a Lagrangian formalism with a Chern-Simons-like term. A
Hamiltonian analysis is also done, what leads  to the quantization
of the system.
\end{abstract}

\maketitle

The study  of  dissipative system in quantum theory is of strong
theoretical interest and of great relevance in practical
applications. However, the standard quantization scheme, that is
based on the existence of either a Hamiltonian or a Lagrangian
function for the system in which we are interested, is not
applicable when the  Lagrangian or Hamiltonian  has
explicit time dependence.

Among some approaches to study dissipative systems, there is one,
where in order to implement a canonical quantization scheme, one
must first double the phase-space dimensions, so as to deal with
an effective isolated system (indirect
representation)\cite{RB,01}. The new degrees of freedom thus
introduced may be  represented by a single equivalent
(collective) degree of freedom for the bath, which absorbs  the
energy dissipated by the system. An important system with
appropriated characteristic that allows one to use the indirect
representation is the study of quantum dynamics of accelerated
charge. It is a dissipative system once an accelerated charge
loses energy, linear momentum, and angular momentum carried by the
radiation field\cite{PRE}. The effect of these losses to the motion of
charge is know as radiation damping\cite{03}.

The process of radiation damping is important in many areas of
electron accelerator operation\cite{04}, like in recent
experiments with intense-laser relativistic-electron scattering at
lasers frequencies and field strengths where radiation reaction
forces begin to become significant \cite{13,14}.

The purpose of this letter is to present a new method of
canonical quantization  of the RD based on the doubling of the
degrees of freedom. This letter is organized as follows. The
Lagrangian and Hamiltonian formalism is derived in the hyperbolic
coordinates, writing down corresponding constraints and their Dirac
Brackets of the independent canonical variables. The Galilean
symmetry algebra has also been provided. Our model can be
described either in terms of phase space variables with commuting
space coordinates, or in terms of new phase space variables with
noncommutative space coordinates. For the choice of phase space
with noncommutative space coordinates, we see that the Hamiltonian
of the model describes a free motion in this space supplemented by
internal degrees of freedom. After considering free motion in the
noncommutative space we introduce interactions in the classical
space with a potential term which depends on noncommuting $D=2$
space coordinates. Introducing the standard quantized oscillator
variables the quantum Hamiltonian has been provided in terms of
the Casimir operator. We also observe that, in the limit where the
dissipation is not present our Hamiltonian formulation describes
the dynamics of the two-undamped harmonic oscillator motion.

We begin  with a review of the problem of the radiation damping.
The equation of motion of the one-dimensional radiation
damping \cite{03}, without external force, is
\be\label{01} m\ddot x -\gamma \stackrel{\ldots}x =0, \ee
where $\gamma ={2\over3}{{me^2}\over {c^3}}$ and $m$  are
independent of time.

Since the system ({\ref{01}) is dissipative a straight-forward
Lagrangian description leading to a consistent canonical formalism is not
available. To develop a canonical formalism we require to consider
(\ref{01}) along with its time-reversed image \cite{01}
\be\label{02} m\ddot y + \gamma \stackrel{\ldots}y =0 \ee
so that the composite system is conservative. The system
(\ref{01}) and (\ref{02}) can be derived from the Lagrangian
\cite{Eu}
\be\label{03} L=m\dot x\dot y +{{\gamma}\over 2}\left( \dot x
\ddot y -\ddot x \dot y \right) \ee
where  $x$ is the RD coordinate and $y$ corresponds to the
time-reversed counterpart. So, the system made of the RD and of
its time-reversed image globally behaves as a closed system.
Introducing the hyperbolic coordinates $x_1$ and $x_2$ \cite{05}
where,
\be\label{04} x = {1\over{\sqrt{2}}}\left(x_1 +x_2 \right); \;\; y
={1\over{\sqrt{2}}}\left(x_1 -x_2 \right) \ee
the above Lagrangian can be written in a compact notation as
\be\label{05} L={m\over 2}g_{ij}\; \dot x_i \dot x_j -{\gamma
\over 2}\epsilon_{ij}\; \dot x_i \ddot x_j \ee
where the pseudo-euclidian metric $g_{ij}$ is given by
$g_{11}=-g_{22}=1$, $g_{12}=0$ and $\epsilon_{12}=-\epsilon_{21}
=1$. This Lagrangian is similar to the one discussed by Lukierski
at all \cite{06}, but in this case we have a pseudo-euclidian
metric. The  equations of motion corresponding to the Lagrangian
(\ref{05}) are
\be\label{05.1} m\ddot x_1 - \gamma \stackrel{\ldots}x_2 =0,\;\;
m\ddot x_2 -\gamma \stackrel{\ldots}x_1 =0. \ee

Now, due to the presence of a second order derivative in the
Lagrangian we have to introduce two momenta
\be\label{06} p_i = {{\partial L}\over{\partial \dot x_i}}
-{d\over{dt}}{{\partial L}\over{\partial \ddot x_i}}\; ;
\;\;\tilde p_i ={{\partial L}\over{\partial \ddot x_i}}. \ee
In our case:
\be\label{07}p_i =mg_{ij} \dot x_j -\gamma \epsilon_{ij} \ddot x_j
\; ;\;\; \tilde p_i ={\gamma\over 2}\epsilon_{ij}\dot x_j. \ee

The Hamiltonian hence reads
\ba\label{08} H&=&\dot x_i p_i + \ddot x_i \tilde p_i
-L\nonumber\\
&=&{m\over 2} g_{ij} \dot x_i \dot x_j -\gamma\epsilon_{ij} \dot
x_i \ddot x_j , \ea
or, using Eq.(\ref{07}), we have
\be\label{09} H={{2m}\over{\gamma^2}}  \tilde p_i g_{ij} \tilde
p_j -{2\over \gamma}p_i \epsilon_{ij}\tilde p_j \ee

Note that, this theory has two constraints,
\be\label{10} \chi_i =\dot x_i +{2\over \gamma}\epsilon_{ij}\tilde
p_j ,\ee
where the eight-dimensional phase space is given by $( x_i , \dot
x_i ,p_i , \tilde p_i )$. These constraints lead to the
replacement  of the canonical Poisson brackets
\be\label{11} \{ x_i, p_j\}=\delta_{ij} \;\;\; \{\dot x_i ,
\tilde
p_j \} =\delta _{ij},
\ee
where the remaining Poisson brackets are all null,
by the Dirac brackets of the independent canonical variables $\(
x_i ,p_i ,\tilde p_i \)$
\be\label{12} \{x_i ,p_i \}_D =\delta_{ij} \;\;\; \{\tilde p_i ,
\tilde p_j \}_D = {\gamma \over 4}\epsilon_{ij},\ee
with all Dirac brackets null. 

We can also use the Faddeev-Jackiw method \cite{08} to obtain the brackets (\ref{12}), introducing a Lagrange multiplier which equates $\dot x$ to $z$, and replace all differentiated $x$-variables in the Lagrangian (\ref{05}) by $z$ variables. Then one has a first-order Lagrangian whose canonical structure can be analyzed by the Faddeev-Jackiw method, yielding the brackets (\ref{12}).

It is important to point out that this model, described by the
Lagrangian, Eq.(\ref{05}), presents Galileo symmetry. In
particular, the  symmetry generators can be expressed as follows

\noindent(i)Translations \be\label{13} P_i =p_i \ee
(ii) Rotations \be \label{14} J=x_i \epsilon_{ij} p_j -{2\over
\gamma}{\tilde p_j }^2 \ee
(iii) Galilean Boosts \be \label{15} \tilde B_i =p_i t -K_i \ee
where \be\label{16}
 K_i = mg_{ij} x_j -2 \tilde p_i .
\ee
The action of these transformations on the model is
straightforward: the time and space arguments are shifted or the
space argument is rotated. Slightly less trivial is the action of
Galileo boosts.

Now, in analogy with what  is done in ref.\cite{06},  our model
can be described in terms of new phase space variables with
noncommutative space coordinates given by relations $\[ X_i ,X_j
\] =i{k\over m^2 }\epsilon_{ij}$ \cite{06,07}.  That is important, because so we can see that the dynamics in the model considered (Eq.(\ref{09})) can be separeted into two independent sectors - describing external and internal dynamics. For our Galilean
system, described by Lagrangian (\ref{05}), the noncommuting
position variables $X_i$ can be expressed, using the Galilean
boosts operator $K_i$, Eq.(\ref{16}), as
\be \label{17}
X_i =x_i -{2\over m}g_{ij}\tilde p_j .\ee
Considering $P_i =p_i$ and redefining the second pair of momenta $\tilde p_i$ as
\be \label{17.1}
\tilde P_i ={\gamma \over {2m}} g_{ij}p_j +\epsilon_{ij}\tilde p_j ,
\ee
we obtain the following  standard canonical Dirac brackets for the
six phase space variables $(X_i ,P_i ,\tilde P_i )$
\ba\label{18}\{X_i ,X_j \}_D  &=&-{\gamma \over
m^2}\epsilon_{ij}\:,\;\{\tilde P_i ,\tilde P_j \}_D ={\gamma \over
4}\epsilon_{ij}, \nonumber\\
 \{X_i , P_j \}_D &=&\delta_{ij} ,\;\;\{X_i , \tilde
P_j \}_D =0, \ea
where the relations given in (\ref{12}) were used.

Note that due to the parameter $\gamma$, noncommutativity is introduced
 in the coordinate sector\cite{08,09,10}. One can
now consider the dynamics of the model using this  noncommutative framework
, and rewrite the Hamiltonian Eq.(\ref{09}), as
\be \label{20} H= {1\over {2m}}P_i g_{ij} P_j -{{2m}\over
\gamma^2} \tilde P_i g_{ij} \tilde P_j . \ee
We thus obtain, in the noncommutative phase space, that the
Hamiltonian, Eq.(\ref{09}), can be diagonalized so that it
describes a free motion (external modes) supplemented by the
oscillator modes (``internal'' modes) with negative sign of their
energies \cite{06}. But, such as in the Ref.\cite{06}, the
variables $\tilde P_i$ can be identified with a standard pair of
canonical variables. Indeed, identifying $\tilde P_1=
\sqrt{\gamma}\;\tilde x$, $\tilde P_2 =\sqrt{\gamma}\;\tilde p$
and introducing oscillator variables
\be\label{20.1} C={1\over \sqrt{2}}(\tilde x +i \tilde p), \;\;
C^* ={1\over \sqrt{2}}(\tilde x -i\tilde p) \ee
we find that the Hamiltonian (\ref{20}) can be rewritten as
\be \label{21} H={1\over {2m}}P_i g_{ij} P_j -{{2m}\over{\gamma}}(
C^2 + {C^*}^2 ) \ee
and  from (\ref{20.1}) that $ \{ C,C^* \}_D =-{i/2}$.

Note that,  not like in Ref.\cite{06}, in our model we
do not need to impose a subsidiary condition$(C|phys\rangle=0)$
because in this case $\langle phys|(C^2 + {C^*}^2)|phys\rangle =0$.
So, we see that the second term in (\ref{21}) do not contribute,
in average, to the spectrum of the Hamiltonian (\ref{21}).

Next we shall introduce interactions to the free Lagrangian
(\ref{05}), a potential energy term, which do not modify the
internal Hamiltonian (second term in (\ref{20})) and add to the
free external Hamiltonian (first term in (\ref{20})) an arbitrary
potential $U(X)$ involving noncommutative variables, as
\be\label{23} H^{(ext)} ={1\over {2m}} P_i g_{ij}P_j + U(X). \ee
It leads to deformation of the constraint algebra, since the
secondary constraint, Eq.(\ref{10}), involves now derivative of the
potential. In the simplest case one can assume that the potential
$U$ is quadratic (electric harmonic potential), so
\be\label{24}  H^{(ext)} = H^{(ext)}_1 - H^{(ext)}_2,\ee
where
\be\label{24.1} H_i^{(ext)}={1\over{2m}}P_i^2 +{{m\omega^2}\over
2}X_i^2, \; i=1,2.\ee
Here $X$ is given by Eq.(\ref{17}) and $\omega$ is the frequency.
Introducing, in the standard way, the oscillator variables
\ba\label{25} A_i &=& \sqrt{{m\omega}\over 2}X_i
+i\sqrt{{1\over{2m\omega}}}P_i \\
A_i^* &=& \sqrt{{m\omega}\over 2}X_i
-i\sqrt{{1\over{2m\omega}}}P_i \ea
we get
\be\label{26} H^{(ext)}_{1(2)} ={\omega \over
2}\(A_{1(2)}A_{1(2)}^* +A_{1(2)}^* A_{1(2)} \). \ee
From the Dirac brackets (see Eq.(\ref{18})) between the
noncommutative variables $X_i$ and $P_i$, using the substitution
$\{\cdot, \cdot\}_D \rightarrow (1/i\hbar )[\cdot,\cdot ]$, one can find that
\ba\label{27} [A_i ,A_j^{\dag} ] &=& \hbar \delta_{ij}
-{{i\hbar\gamma
\omega}\over{2m}}\epsilon_{ij} \;,\\
\[ A_i ,A_{j} \] &=& [A_i^{\dag} ,A_j^{\dag} ] =-{{i\hbar\gamma
\omega}\over{2m}}\epsilon_{ij} \;. \ea
The parameter $\gamma$ introduces a deformation of the Heisenberg
commutation relations, what do not obstruct the quantization  of
the model as well discussed by Banerjee et al in \cite{RB}.
However, we are interested to quantize the model in the
commutative phase space. To this end we build a commutative phase
space introducing the following space coordinates
\ba\label{28} \hat X_i &=&X_i -{\gamma\over{2m^2}}
\epsilon_{ij}P_j \nonumber\\&=&x_i -{2\over m} g_{ij} \tilde p_j
-{\gamma\over{2m^2}} \epsilon_{ij}p_j, \ea
where
\ba\label{29} \{\hat X_i ,P_j \}_D &=&\delta_{ij},\nonumber\\
\{\hat X_i ,\hat X_j \}_D &=& \{\hat X_i ,\tilde P_j \}_D =0 . \ea

Hence we can write the Hamiltonian (\ref{24}) using variables
$(\hat X ,P)$\cite{07} as follows
\be\label{30} H_i^{(ext)} ={P_i^2\over{2\tilde m}}+ {{\tilde m
\tilde \omega^2}\over 2}\hat X_i^2 +{{\gamma
\omega^2}\over{2m}}\epsilon_{ij} \hat X_i P_j , \ee
where $i=1,2$ and
\ba\label{31} \tilde m &=& m\(1+\omega^2 {\gamma^2\over {4m^2}}
\)^{-1}, \\
\tilde \omega^2 &=&\omega^2 \(1+ \omega^2 {\gamma^2 \over
{4m^2}}\). \ea
If we introduce the standard quantized oscillator variables
\ba\label{32} a_i &=& \sqrt{{{\tilde m\tilde
\omega}\over{2\hbar}}} \hat X_i +i\sqrt{{1\over{2\tilde m\tilde
\omega \hbar}}}P_i ,\\
a_i^{\dag} &=&\sqrt{{{\tilde m\tilde \omega}\over{2\hbar}}} \hat
X_i -i\sqrt{{1\over{2\tilde m\tilde \omega \hbar}}}P_i ,\ea
we find that
\be\label{33} H^{(ext)} =\hbar \tilde \omega (a_1^{\dag} a_1 -
a_2^{\dag}a_2 ) + i\hbar {{\omega^2 \gamma}\over {2m}} (a_1^{\dag}
a_2^{\dag} -a_1 a_2 ). \ee
Introducing the following notation:
\be\label{34} \Omega=\omega \( 1+ \omega^2 {\gamma^2 \over {4m^2}}
\)^{1/2} , \;\; \Gamma =\omega^2 {\gamma\over{2m}},\ee
the Eq.(\ref{33}) can be rewritten as
\be\label{35} {\cal H}^{(ext)} = {\cal H}_0 +{\cal H}_I \ee
where
\be \label{36} {\cal H}_0 =\hbar \Omega (a_1^{\dag} a_1 -
a_2^{\dag}a_2 ), \ee
and
\be\label{37} {\cal H}_I = i\hbar \Gamma (a_1^{\dag} a_2^{\dag}
-a_1 a_2 ).\ee

Following the  Ref.\cite{12}, we can see that the dynamical group
structure associated with our system  is that of $SU(1,1)$. The
generators of this algebra are
\ba\label{38} J_{+}&=&a_1^{\dag}a_2^{\dag} ,\;\; J_{-}
=J_{+}^{\dag} =a_1 a_2 ,\\
J_3 &=&{1\over 2}(a_1^{\dag}a_1 +a_2^{\dag} a_2 + 1), \ea
corresponding to the Casimir operator  ${\cal C} ={1\over 4}
+J_3^2 -{1\over 2} (J_+ J_- + J_- J_+ ) ={1\over 4}(a_1^{\dag} a_1
- a_2^{\dag} a_2 )^2 .$
The Hamiltonian (\ref{36}) and (\ref{37}) are then rewritten as
\be \label{40} {\cal H}_0 = 2\hbar \Omega {\cal C},\;\;\; {\cal
H}_I =- 2\hbar \Gamma J_2, \ee
where $\[ {\cal H}_0 ,\; {\cal H}_I \]=0$,
as ${\cal H}_0$ is in the center of the dynamical algebra.

Let us denote by ${\left|n_1 ,n_2 \right\rangle }$ the set of simultaneous eigenvectors of $a_1^{\dag}a_1$ and $a_2^{\dag}a_2$, with $n_1$, $n_2$ non-negative integers. One can see that the eigenvalue of ${\cal H}_0$ in this frame is the constant quantity 2$\hbar\Omega(n_1 - n_2 )$. The eigenstates of ${\cal H}_I$ can be written in the standard basics, in terms of the eigenstates of $(J_3 -{1\over 2})$ in the representation labelled by the value $j \in Z_{1/2}$ of $\cal C$, $\{\left|j,m \right\rangle ; m \geq x\left|j\right|\}$:
\be\label{40.1}{\cal C}\left| j,m \right\rangle =j\left|j,m \right\rangle, \;\;\; j={1\over 2}(n_1 - n_2 );
\ee
\be\label{40.2}
(J_3 -{1\over 2})\left|j,m \right\rangle =m\left|j,m\right\rangle ,\;\;\; m={1\over2}(n_1 + n_2 ).
\ee
Note that this Hamiltonian is similar to the one obtained in
Ref.\cite{12} to the damped harmonic oscillator, with a difference
in the notation introduced in (\ref{34}).
It is important to point out that such as in Ref.\cite{12}, our
Hamiltonian formulation (\ref{35}) is the simple undamped harmonic
oscilator when $\gamma \rightarrow 0$ $(\Omega \rightarrow \omega
)$.  The states generates by $a_2^{\dag}$ represent the sink where
the energy dissipated by the accelerated charge particle flows.
We see therefore that $a_2$-system thus represents the reservoir
or heat bath coupled to the $a_1$-system.


In conclusion, we have shown that in the pseudo-Euclidean metrics
the Lagrangian density of the system made of a charge interacting
with its own radiation and of its time-reversed image, this last
introduced by doubling the degrees os freedom as required by
canonical formalism, actually behaves as a closed system described
by the Lagrangian (\ref{03}). On the hyperbolic plane, the
equation (\ref{05.1}) shows that the dissipative term actually
acts as a coupling between the systems $x_1$ and $x_2$.  Our model
can be interpreted as describing a free motion in the $D=2$ space
with noncommuting coordinates and internal modes with negative
energies. However, because of the pseudo-Euclidian metrics, we
have shown that this internal modes do not contribute to the
energy spectrum. We do not need to impose subsidiary condition.
Finally, by introducing the commuting position variables (see
Eq.(\ref{28})), we observe that the quantum  Hamiltonian is
obtained and that the dynamical group structure associated with
our system is that of $SU(1,1)$.  In future works, we will study the supersymmetric extension and the introdution of gauge interactions into the  model.

This work is supported in part by FAPEMIG and CNPq, Brazilian
Research Agencies. In particular, ACRM and WO would like to acknowledge
the CNPq and CN, WO and FIT would like to acknowledge the FAPEMIG.

\end{document}